\documentclass[%
prb, twocolumn, superscriptaddress,
 amsmath,amssymb,
 aps
]{revtex4-2}

\usepackage{graphicx}
\usepackage{dcolumn}
\usepackage{bm}
\usepackage{xcolor}

\usepackage{lipsum}

\begin{document}

\preprint{APS/123-QED}

\title{Vortical versus skyrmionic states in the topological phase of a twisted bilayer with $d$-wave superconducting pairing}

\author{Leonardo R. Cadorim}
\affiliation{Departamento de F\'isica, Faculdade de Ci\^encias,
Universidade Estadual Paulista (UNESP), Caixa Postal 473, 
17033-360, Bauru-SP, Brazil}
\affiliation{Department of Physics, University of Antwerp, Groenenborgerlaan 171, B-2020 Antwerp, Belgium}
\author{Edson Sardella}
\affiliation{Departamento de F\'isica, Faculdade de Ci\^encias,
Universidade Estadual Paulista (UNESP), Caixa Postal 473, 
17033-360, Bauru-SP, Brazil}
\author{Milorad V. Milo{\v{s}}evi{\'c}}%
 \email[Corresponding author: ]{milorad.milosevic@uantwerpen.be}
\affiliation{Department of Physics, University of Antwerp, Groenenborgerlaan 171, B-2020 Antwerp, Belgium}%
\affiliation{NANOlab Center of Excellence, University of Antwerp, Belgium}

\begin{abstract}
It was recently shown that a chiral topological phase emerges from the coupling of two twisted monolayers of superconducting Bi\textsubscript{2}Sr\textsubscript{2}CaCu\textsubscript{2}O\textsubscript{8$+\delta$} for certain twist angles. In this work, we reveal the behavior of such twisted superconducting bilayers with $d_{x^2-y^2}$ pairing symmetry in presence of applied magnetic field. Specifically, we show that the emergent vortex matter can serve as smoking gun for detection of topological superconductivity in such bilayers. Moreover, we report two distinct skyrmionic states that characterize the chiral topological phase, and provide full account of their experimental signatures and their evolution with the twist angle.
\end{abstract}

\maketitle


\section{\label{sec:sec1}Introduction}

Chiral superconductivity \cite{kallin2016} has been a topic of tremendous interest in the recent literature due to its rich phenomenology \cite{sigrist1991,vojta2000}, including appearance of nontrivial surface currents \cite{stone2004} and half-quantum vortices \cite{read2000,garaud2012,jang2011,zyuzin2017,becerra2016}, to name a few examples. Being mostly characterized by several Fermi surfaces, chiral superconductors often present multiple superconducting gaps, and are thereby prone to a plethora of interesting physics typical of multicomponent superconductivity \cite{milosevic2015,tanaka2015,babaev2005,lin2011,tanaka2001}. Arguably, chiral superconductors gained a special relevance due to the increasing interest in topological superconductivity \cite{sato2017} and its promise towards use in modern quantum technologies \cite{sarma2015}. With its highly non-trivial topology, the chiral state of superconductors is known to present the uniquely associated phenomena, such as the gapless edge states \cite{volovik1997} and Majorana bound states localized in the vortex cores \cite{volovik1999}, which obey the non-Abelian statistics \cite{ivanov2001} fundamental to future applications in quantum computing.

Recently, Can \textit{et al.} \cite{can2021} showed that a twisted bilayer composed of two monolayers of the high-temperature superconductor Bi\textsubscript{2}Sr\textsubscript{2}CaCu\textsubscript{2}O\textsubscript{8$+\delta$} \cite{yu2019} (Bi-2212) can display a chiral topological phase which breaks time-reversal symmetry for twist angles near $45^{\circ}$. As they argued, at a twist angle equal to $45^{\circ}$, the $d_{x^2-y^2}$ order parameter of each layer, characteristic of Bi-2212, induces a significant $d_{xy}$ component in the order parameter of the other layer. This results in a superconducting state with $d+id'$ pairing symmetry.

The above arguments were developed in Ref.~\onlinecite{can2021} considering a homogeneous superconducting state. In the present work we go beyond this premise and investigate how such system responds to applied magnetic field, i.e. how the vortex matter of such bilayers evolves with the twist angle between the monolayers. As we will show, the emergent typical vortex configurations can be used as a smoking gun for the detection of the chiral topological phase. Namely, due to the broken time-reversal symmetry, skyrmionic vortex states \cite{babaev2002b,becerra2016,garaud2013,benfenati2022} arise in the topological phase, and are clearly identifiable by their magnetic signature. Moreover, we show that the vortex matter changes even \textit{within} the topological phase itself. Namely, as one varies the twist angle in the range where topological phase is stable, two different skyrmionic states are found. In one of them, states with unit topological charge are favored (presenting as a lattice of vortex pairs), while in the other one states with large topological charge become energetically favorable, causing formation of extended vortex chains with a distinct appearance and magnetic signature.

The outline of this work is as follows. In Sec.~\ref{sec:sec2} we present our theoretical formalism and show how we deal with the twisted bilayer system at hand within the framework of the Ginzburg-Landau theory. In Sec.~\ref{sec:sec3} we present and discuss our main results. We start by showing the existence of a topological phase for certain values of the twist angle in a homogeneous system, to subsequently reveal and characterize the vortical and skyrmionic matter, as well as transitions between them, inside the topological phase. Our concluding remarks are given in Sec.~\ref{sec:sec4}.


\section{\label{sec:sec2}The theoretical model}

The free energy density of our system can be described as a sum of three parts $\mathcal{F} = F_1+F_2+F_{12}$, with
\begin{eqnarray}
    F_1 &=& -2\alpha_s|\Delta_{(s1)}|^2-|\Delta_{(d1)}|^2+\frac{4}{3}|\Delta_{(s1)}|^4+\frac{1}{2}|\Delta_{(d1)}|^4 \nonumber \\
    &&+\frac{8}{3}|\Delta_{(s1)}|^2|\Delta_{(d1)}|^2+\frac{2}{3}(\Delta^{*2}_{(s1)}\Delta^2_{(d1)}+H.c.) \nonumber \\
    &&+2|\bm{\Pi}\Delta^* _{(s1)}|^2+|\bm{\Pi}\Delta^*_{(d1)}|^2+(\Pi_x\Delta^*_{(s1)}\Pi^*_x\Delta_{(d1)} \nonumber \\
    &&-\Pi_y\Delta^*_{(s1)}\Pi^*_y\Delta_{(d1)}+H.c.)
    \label{eq:eq1}
\end{eqnarray}
being the free energy of the unrotated (non-twisted) layer. Here, $\Delta_{(s1)}$ and $\Delta_{(d1)}$ are the order parameters corresponding to the $s$ and $d$-wave pairings, respectively \cite{zhang2020}. Hereafter, the subscript $(1)$ indicates the order parameters of the unrotated layer, while the subscript $(2)$ denotes the rotated (twisted) one. In this work, we add the $s$ component of the order parameter in order to induce the correct fourfold symmetry in the $d$ component. This is done through the mixed gradient terms in the free-energy. The parameter $\alpha_s$ determines the relative strength between the $s$ and $d$ order parameters. Once we are mainly interested in the condensate with $d$-wave pairing, we use $\alpha_s = 0.7$ that leads to a weak modulus for the $s$-wave order parameter (under 20\% of the $d$-wave order parameter). We also define the momentum operator $\bm{\Pi} = i\bm{\nabla}-\bm{A}$ for compacter presentation of the formulae. Here $\bm{A}$ stands for the magnetic vector potential due to the applied magnetic field and the magnetic response of the superconducting layers.

The second contribution to the free energy stems from the rotated layer, and reads
\begin{eqnarray}
    F_2 &=& -2\alpha_s|\Delta_{(s2)}|^2-|\Delta_{(d2)}|^2+\frac{4}{3}|\Delta_{(s2)}|^4+\frac{1}{2}|\Delta_{(d2)}|^4 \nonumber \\
    &&+\frac{8}{3}|\Delta_{(s2)}|^2|\Delta_{(d2)}|^2+\frac{2}{3}(\Delta^{*2}_{(s2)}\Delta^2_{(d2)}+H.c.) \nonumber \\
    &&+2|\bm{\Pi}\Delta^*_{(s2)}|^2+|\bm{\Pi}\Delta^*_{(d2)}|^2+\cos(2\theta)\Pi_x\Delta^*_{(s2)}\Pi^*_x\Delta_{(d2)} \nonumber \\
    &&-\cos(2\theta)\Pi_y\Delta^*_{(s2)}\Pi^*_y\Delta_{(d2)}-\sin(2\theta)\Pi_x\Delta^*_{(s2)}\Pi^*_y\Delta_{(d2)} \nonumber \\
    &&-\sin(2\theta)\Pi_y\Delta^*_{(s2)}\Pi^*_x\Delta_{(d2)}+H.c. .
    \label{eq:eq2}
\end{eqnarray}
Here, $\theta$ is the twist angle and the expression for the mixed gradient terms presented in Eq.~\eqref{eq:eq2} is obtained after transformation on the momentum operator from the rotated coordinates to the unrotated ones.

The final contribution to the free energy captures the interaction between the two layers \cite{can2021}
\begin{eqnarray}
    F_{12} = &A&|\Delta_{(d1)}|^2|\Delta_{(d2)}|^2-B\cos(2\theta)(\Delta_{(d1)}\Delta^*_{(d2)}+H.c.) \nonumber \\
    +&C&(\Delta^2_{(d1)}\Delta^{*2}_{(d2)}+H.c.),
    \label{eq:eq3}
\end{eqnarray}
with $A$, $B$ and $C$ taken as phenomenological (free) parameters. The term proportional to $B$ in Eq.~\eqref{eq:eq3} depends on $\cos(2\theta)$ due to symmetry reasons, as discussed in Ref.~[\onlinecite{can2021}], and represents the tunneling of Cooper pairs between the two layers. Following same reasoning, one can interpret
the term proportional to $C$ as the coherent tunnelling of two Cooper pairs between the layers.

In the above equations, all lengths are expressed in units of the coherence length $\xi = (\nu_F/2)\sqrt{W/\ln(T_d/T)}$, with $W = 7\xi(3)/(8\pi^2T^2)$, the order parameters are in units of $\Delta_0 = \sqrt{(4/3W)\ln(T_d/T)}$, the magnetic field is in units of $H_{c2} = \Phi_0/(2\pi\xi^2)$, where $\Phi_0 = hc/2e$ is the magnetic flux quantum, and the free energy density is in units of $F_0 = (4/3W)\ln(T_d/T)$. For details on the derivation of the free energy for a single layer we refer to Refs.~[\onlinecite{zhang2020},\onlinecite{ren1995}].

Minimizing the total energy $\mathcal{F} = F_1+F_2+F_{12}$ with respect to the order parameters we arrive to the appropriate Ginzburg-Landau equations:
\begin{eqnarray}
    &&-\alpha_s\Delta_{(s1)}+\frac{4}{3}|\Delta_{(s1)}|^2\Delta_{(s1)}+\frac{4}{3}|\Delta_{(d1)}|^2\Delta_{(s1)}+\frac{2}{3}\Delta^2_{(d1)}\Delta^*_{(s1)} \nonumber \\
    &&+\Pi^{*2}\Delta_{(s1)}+\frac{1}{2}(\Pi^{*2}_x-\Pi^{*2}_y)\Delta_{(d1)} = 0,
    \label{eq:eq4}
\end{eqnarray}
\begin{eqnarray}
    &&-\Delta_{(d1)}+|\Delta_{(d1)}|^2\Delta_{(d1)}+\frac{8}{3}|\Delta_{(s1)}|^2\Delta_{(d1)}+\frac{4}{3}\Delta^2_{(s1)}\Delta^*_{(d1)} \nonumber \\
    &&+A|\Delta_{(d2)}|^2\Delta_{(d1)}-B\cos(2\theta)\Delta_{(d2)}+2C\Delta^2_{(d2)}\Delta^*_{(d1)} \nonumber \\
    &&+\Pi^{*2}\Delta_{(d1)}+(\Pi^{*2}_x-\Pi^{*2}_y)\Delta_{(s1)} = 0,
    \label{eq:eq5}
\end{eqnarray}
\begin{eqnarray}
    &&-\alpha_s\Delta_{(s2)}+\frac{4}{3}|\Delta_{(s2)}|^2\Delta_{(s2)}+\frac{4}{3}|\Delta_{(d2)}|^2\Delta_{(s2)} \nonumber \\
    &&+\frac{2}{3}\Delta^2_{(d2)}\Delta^*_{(s2)}+\frac{1}{2}(\cos(2\theta)\Pi^{*2}_x-\cos(2\theta)\Pi^{*2}_y)\Delta_{(d2)} \nonumber \\ &&-\frac{1}{2}\sin(2\theta)\Pi^*_x\Pi^*_y\Delta_{(d2)}-\frac{1}{2}\sin(2\theta)\Pi^*_y\Pi^*_x\Delta_{(d2)} \nonumber \\
    && +\Pi^{*2}\Delta_{(s2)} = 0,
    \label{eq:eq6}
\end{eqnarray}
and
\begin{eqnarray}
    &&-\Delta_{(d2)}+|\Delta_{(d2)}|^2\Delta_{(d2)}+\frac{8}{3}|\Delta_{(s2)}|^2\Delta_{(d2)}+\frac{4}{3}\Delta^2_{(s2)}\Delta^*_{(d2)} \nonumber \\
    &&+A|\Delta_{(d1)}|^2\Delta_{(d2)}-B\cos(2\theta)\Delta_{(d1)}+2C\Delta^2_{(d1)}\Delta^*_{(d2)} \nonumber \\
    &&+\Pi^{*2}\Delta_{(d2)}+(\cos(2\theta)\Pi^{*2}_x-\cos(2\theta)\Pi^{*2}_y)\Delta_{(s2)} \nonumber \\
    &&-\sin(2\theta)\Pi^*_x\Pi^*_y\Delta_{(s2)}-\sin(2\theta)\Pi^*_y\Pi^*_x\Delta_{(s2)}= 0.
    \label{eq:eq7}
\end{eqnarray}
Eqs. \eqref{eq:eq4}-\eqref{eq:eq7} are then solved for different twist angles $\theta$, assuming periodic boundary conditions \cite{doria1989}. As the Ginzburg-Landau parameter $\kappa$ for Bi-2212 is typically much greater than 1, we disregard the contribution of the supercurrents to the total magnetic field and use a vector potential solely due to the applied magnetic field to solve the above set of equations.

For a given $\theta$, we initialize the calculations from dozens of different initial conditions for the order parameters, while also varying the aspect ratio of the unit cell of the simulation in order to identify the lowest-energy solutions for the vortex states. In what follows, we display the ground-state found for an external applied magnetic field that corresponds to the flux of $24\Phi_0$ threading the shown unit cell, without loss of generality. Namely, for other values of the applied magnetic field we obtained qualitatively equivalent results.

\section{\label{sec:sec3}Results and Discussion}

As shown in Ref.~[\onlinecite{can2021}] in the case of homogeneous superconductivity, for a certain range of $\theta$ the competition between the terms proportional to $B$ and $C$ in $F_{12}$ yields a non-trivial phase difference between the $d$-wave components of the order parameters of the two layers. Let us start by discussing the homogeneous solutions of our free-energy model and show that it analogously allows for the existence of a topological phase.

\begin{figure}[!t]
    \centering
    \includegraphics[width=\columnwidth]{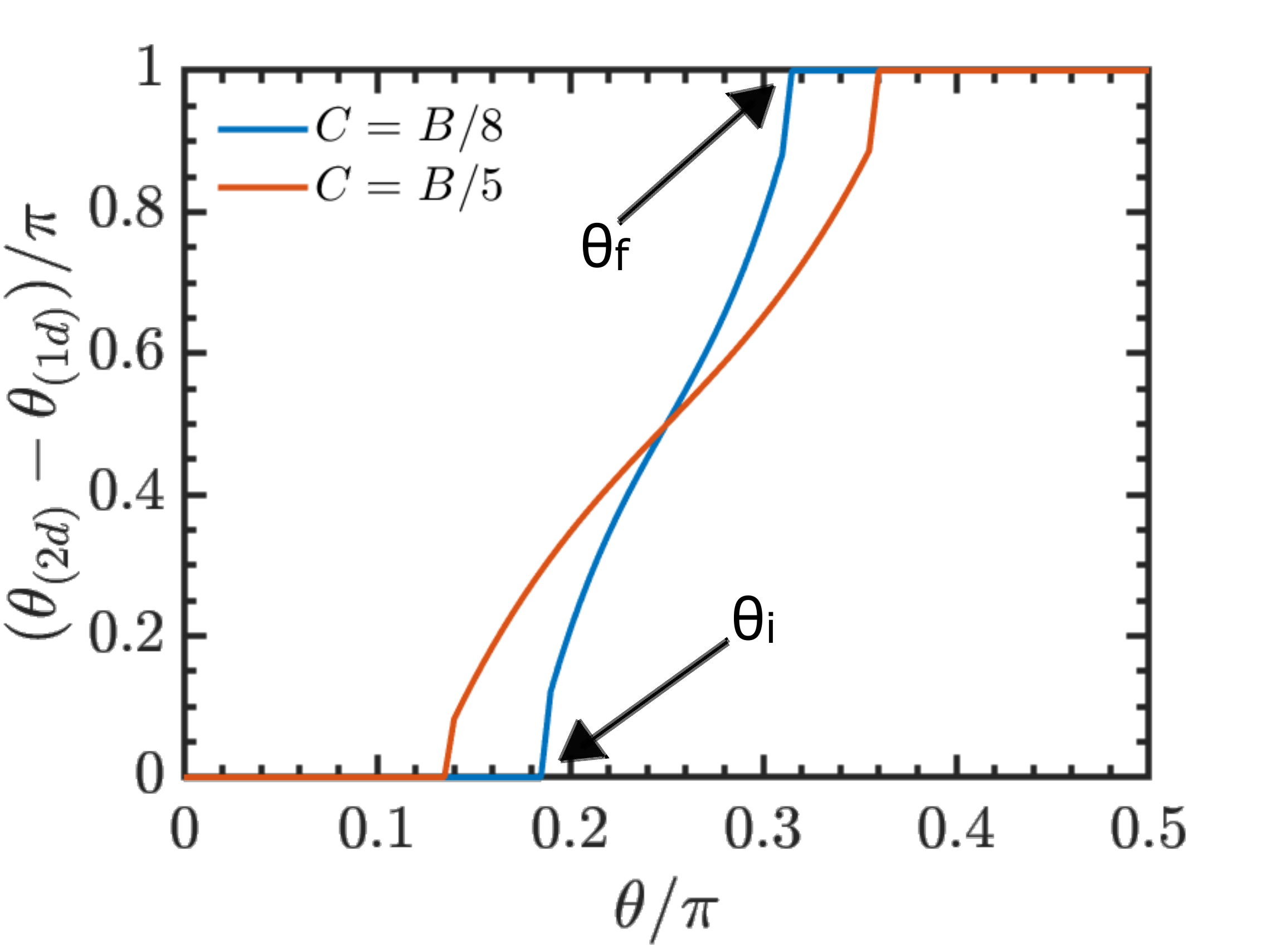}
    \caption{(Color online) Phase difference between the $d$-wave components of the order parameters of two layers as a function of the twist angle $\theta$. Blue and red curves represent the phase difference for $C = B/8$ and $C = B/5$, respectively, while $A = B = 0.1$. The nontrivial values of the phase difference ($\neq 0$ or $\pi$) indicate existence of a topological phase for a particular twist angle.}
    \label{fig:fig1}
\end{figure}

To do this, we minimize the free energy density $\mathcal{F} = F_1+F_2+F_{12}$ with respect to the modulus and phase of the $s$ and $d$ components of the order parameter in both layers. Fig.~\ref{fig:fig1} shows the phase difference between $\Delta_{(d1)}$ and $\Delta_{(d2)}$ which minimizes the free energy as a function of $\theta$. As can be seen from the figure, for small twist angles up to a critical angle $\theta_i$, the free energy is minimal when the order parameters have the same phase. For twist angles larger than a critical value $\theta_f$, the phase difference that yields minimal energy equals $\pi$.

On the other hand, for angles between $\theta_i$ and $\theta_f$, one obtains a non-trivial phase difference between the condensates of the two layers, which means a superconducting state that breaks time-reversal symmetry. In particular, for $\theta = \pi/4$, the phase difference is equal to $\pi/2$, i.e. a $d+id'$ superconducting state is found \cite{can2021}. The values of $\theta_i$ and $\theta_f$ depend on the particular values chosen for the parameters $A$, $B$ and $C$, as can be seen from the two examples shown in Fig.~\ref{fig:fig1}. Nevertheless, the features of the superconducting state that we discuss below are always present in the range $\theta_i < \theta < \theta_f$, for any choice of the aforementioned parameters. Therefore, without loss of generality of our results, in what follows we will use the parameters correspondent to the blue curve in Fig.~\ref{fig:fig1}. In that case, $\theta_i \approx 34^{\circ}$ and $\theta_f \approx 56^{\circ}$.
\begin{figure}[!t]
    \centering
    \includegraphics[width=\columnwidth]{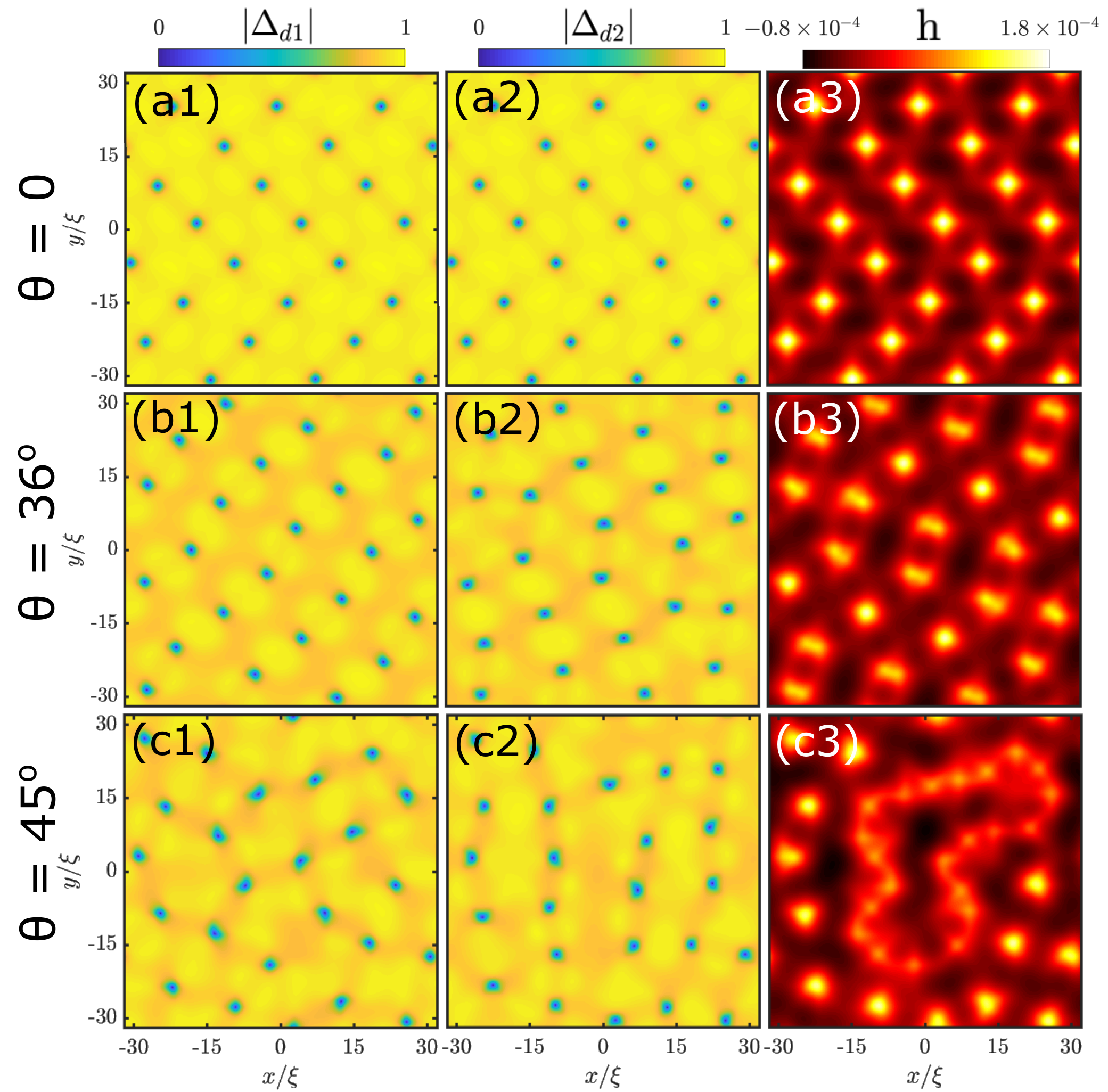}
    \caption{(Color online) Vortex configurations in the $d$-wave component of the order parameter of the unrotated layer (first column) and the rotated layer (second column), and the magnetic response of the system (third column), for three selected twist angles.}
    \label{fig:fig2}
\end{figure}

\begin{figure*}[!t]
    \centering
    \includegraphics[width=\textwidth]{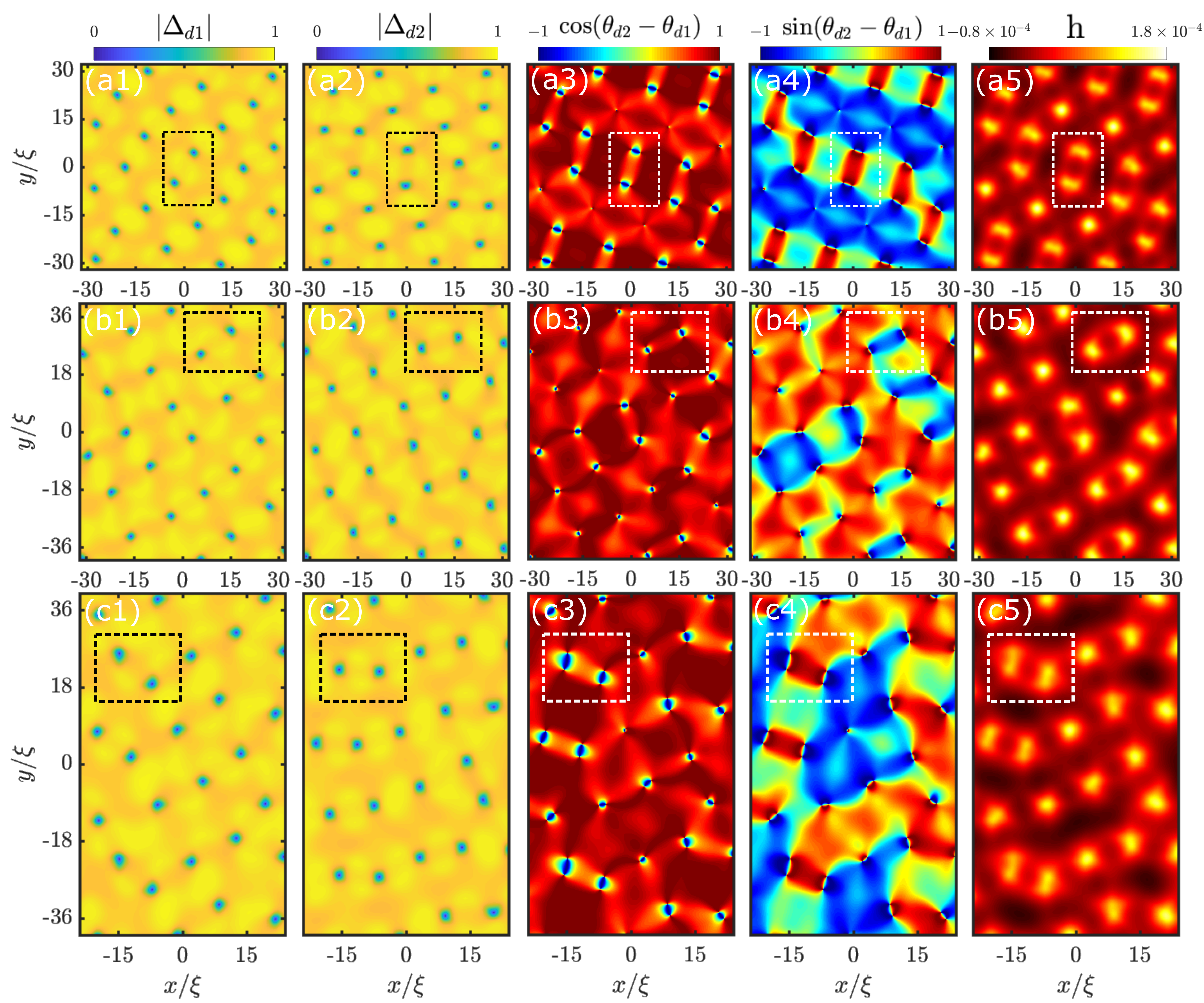}
    \caption{(Color online) Vortex configurations for twist angle $\theta = 36^{\circ}\gtrsim\theta_i$ at three values of applied magnetic field. Each row corresponds to a different system size (lateral sizes are shown), for fixed magnetic flux of $24\Phi_0$. From left to right, the columns respectively show the magnitude of the $d$-wave component of the order parameter for unrotated and rotated layers, the cosine and sine of the phase difference between the order parameters in two layers, and the magnetic field distribution across the system.}
    \label{fig:fig3}
\end{figure*}

Let us now go beyond these results and show how the twist angle affects the vortex matter of such bilayers. In Fig.~\ref{fig:fig2} we show the spatial distribution of the $d$-wave component of the order parameter for both layers, together with the magnetic field distribution in the system, for $\theta = 0$, $36^{\circ}$ and $45^{\circ}$. When plotting the magnetic field profile, we consider only the contribution of the supercurrents, after subtracting the (strongly dominating) homogeneous external field from the total one. For $\theta = 0$, as discussed above, the phase difference between the order parameters is locked at zero. It is therefore energetically favorable for the vortices in two layers to organize in a composite configuration, where the vortex cores are vertically aligned between the unrotated and the rotated layer. In this case, the magnetic field profile of the vortices exhibits well defined peaks at vortex locations and one can clearly distinguish the fourfold symmetry characteristic of $d_{x^2-y^2}$ superconductors. The vortex configurations remain qualitatively unchanged for the non-zero twist angles outside the topological phase, i.e. for $\theta < \theta_i$ or $\theta > \theta_f$.

Next we increase the twist angle to $36^{\circ}$, larger than $\theta_i$, i.e. upon entry to the topological phase in the ground state of the system. The composite vortex configuration is no longer the most energetically favorable state, as the vortex cores in two layers no longer coincide. As a consequence, the field of the vortex is now distributed over the two displaced cores in two layers (cf. Fig.~\ref{fig:fig1}), reflecting a magnetic field profile of a dimer rather than one clear peak. Notably, some vortices within the configuration remain seemingly composite. The situation radically changes as we increase the twist angle further, to $\theta = 45^{\circ}$ (bottom row of Fig.~\ref{fig:fig1}). Here one sees that not only vortex cores displace between the layers, they also organize into extended closed vortex chains. As will be discussed in the following paragraphs, such a vortex chain is formed along a domain wall separating sample regions with different phase differences between the layers. Moreover, such chains will exhibit skyrmionic topology, with an integer topological charge equal to the total vorticity of the chain. Last but not least, the overall shape and the magnetic signature of these chains are uniquely distinct which facilitates their experimental observation.

These results demonstrate that the twist angle and the onset of a topological phase strongly influence the vortex matter of the system, with detectable consequences in the magnetic profile at the onset of the topological phase and within the topological phase itself. This feature can therefore be used as a smoking gun for the detection of topological superconductivity in such and similar bilayers. In what follows, we further detail the vortex configurations for $\theta = 36^{\circ}$ and $\theta = 45^{\circ}$, which are the representative examples of two different types of behavior we encountered in the vortex matter inside the topological phase.

\begin{figure*}[!t]
    \centering
    \includegraphics[width=\textwidth]{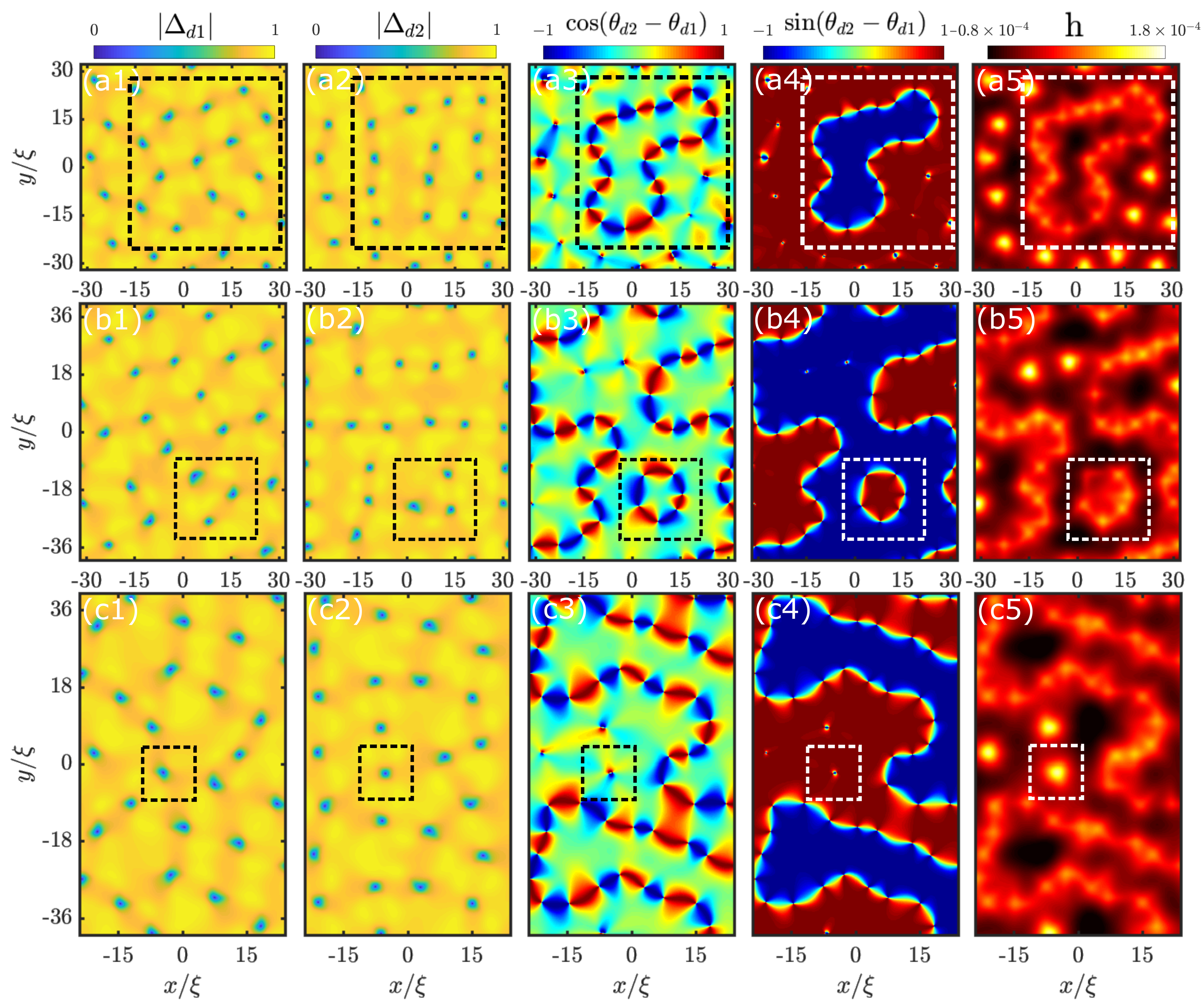}
    \caption{(Color online) Vortex configurations for $\theta = 45^{\circ}$, deep inside the topological phase, for three values of applied magnetic field. Each row corresponds to a different system size (lateral sizes are shown), for fixed magnetic flux of $24\Phi_0$. From left to right, the columns respectively show the magnitude of the $d$-wave component of the order parameter for unrotated and rotated layers, the cosine and sine of the phase difference between the order parameters in two layers, and the magnetic field distribution across the system.}
    \label{fig:fig4}
\end{figure*}

\begin{figure}[!t]
    \centering
    \includegraphics[width=\columnwidth]{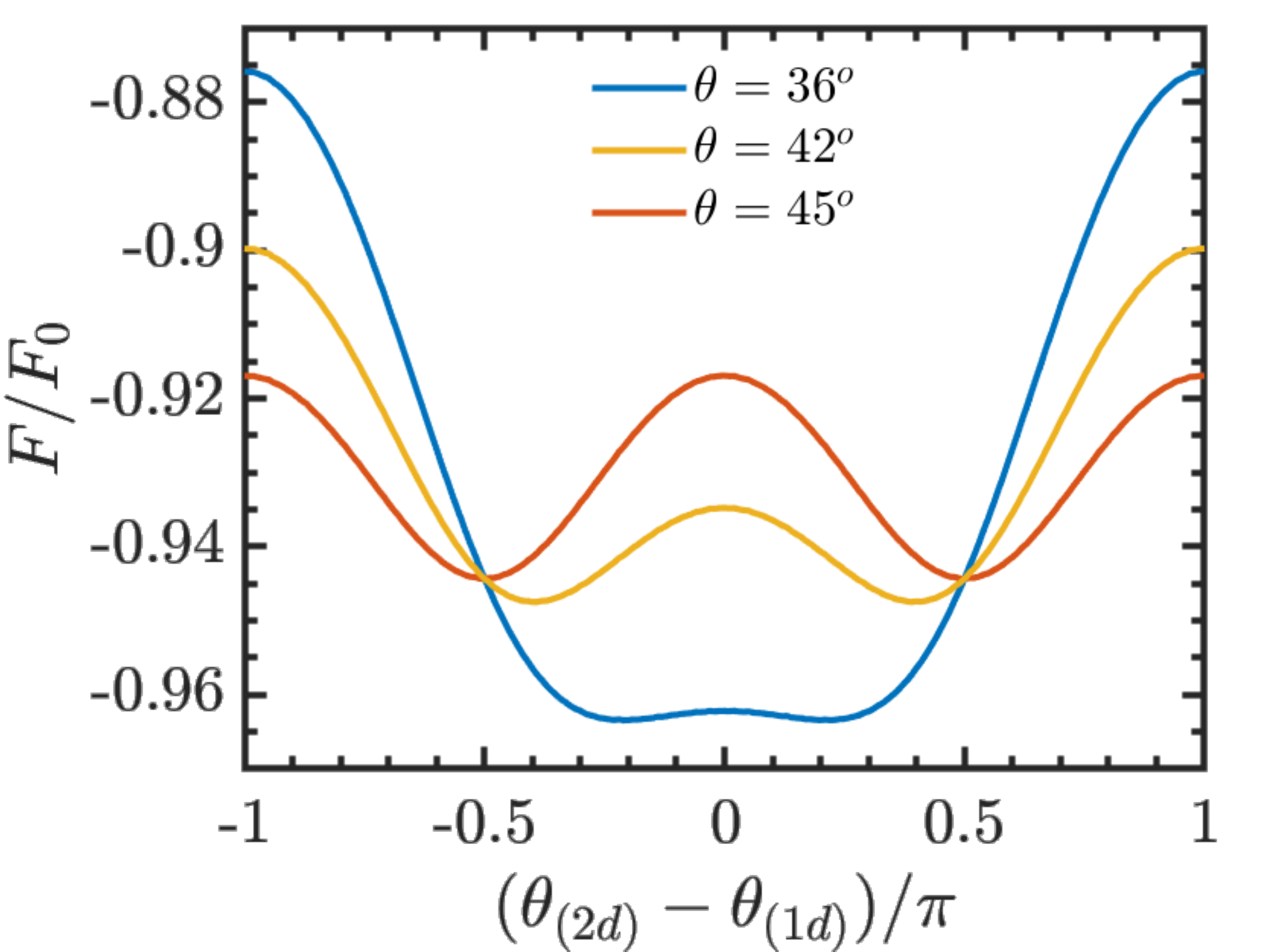}
    \caption{(Color online) Energy of the homogeneous system (without vortices) as a function of the phase difference between the $d$-wave components of the order parameter in two twisted layers. The blue, yellow and red curves show the energy for the twist angle $\theta = 36^{\circ}$, $\theta = 42^{\circ}$ and $\theta = 45^{\circ}$,       respectively.}
    \label{fig:fig5}
\end{figure}

\begin{figure*}[!t]
    \centering
    \includegraphics[width=\textwidth]{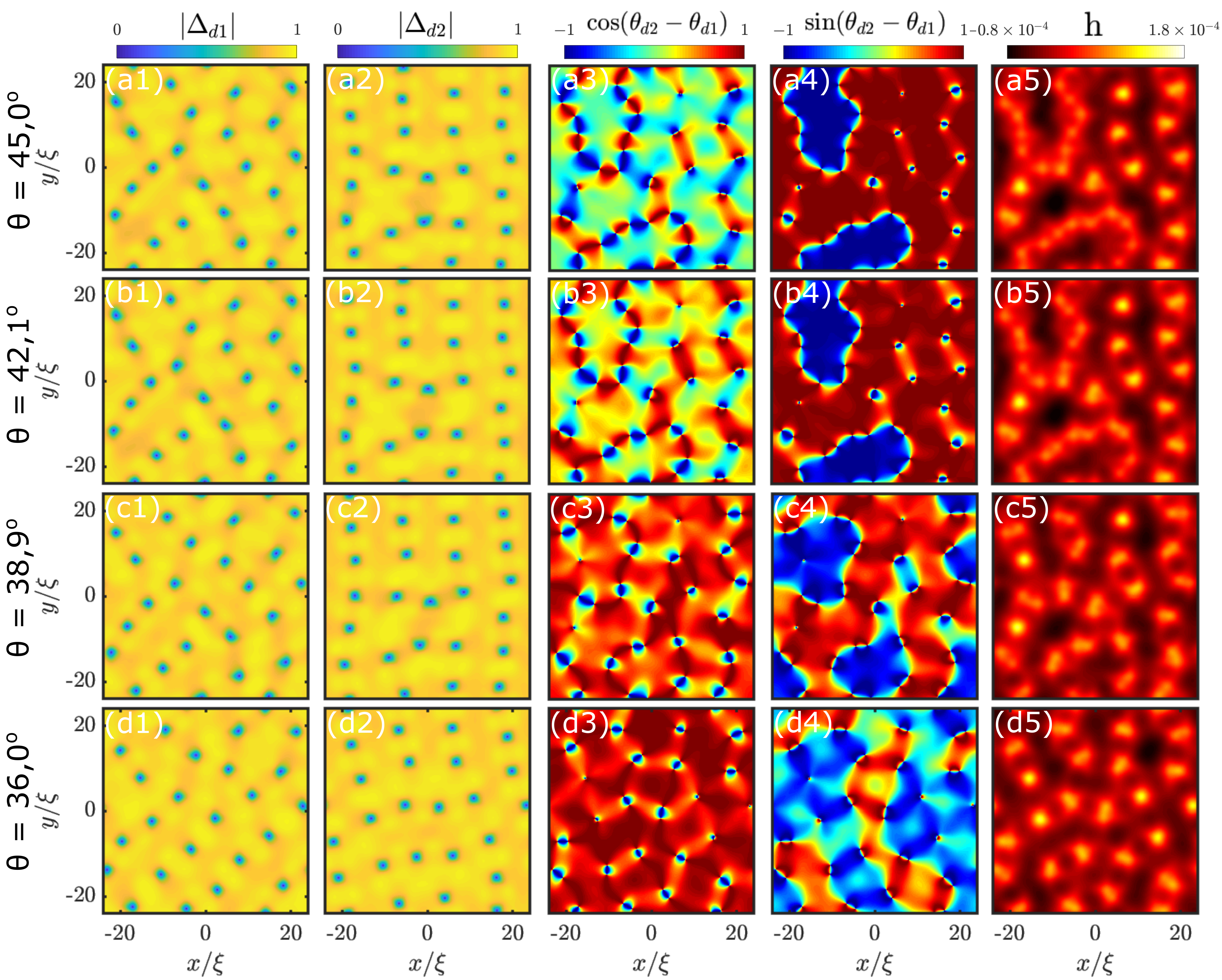}
    \caption{(Color online) Evolution of the vortex configuration when adiabatically decreasing the twist angle from $\theta = 45^{\circ}$ to $36^{\circ}$. From left to right, the columns respectively show the magnitude of the $d$-wave component of the order parameter for unrotated and rotated layers, the cosine and sine of the phase difference between the order parameters in two layers, and the magnetic field distribution across the system.}
    \label{fig:fig6}
\end{figure*}

\begin{figure*}[!t]
    \centering
    \includegraphics[width=\textwidth]{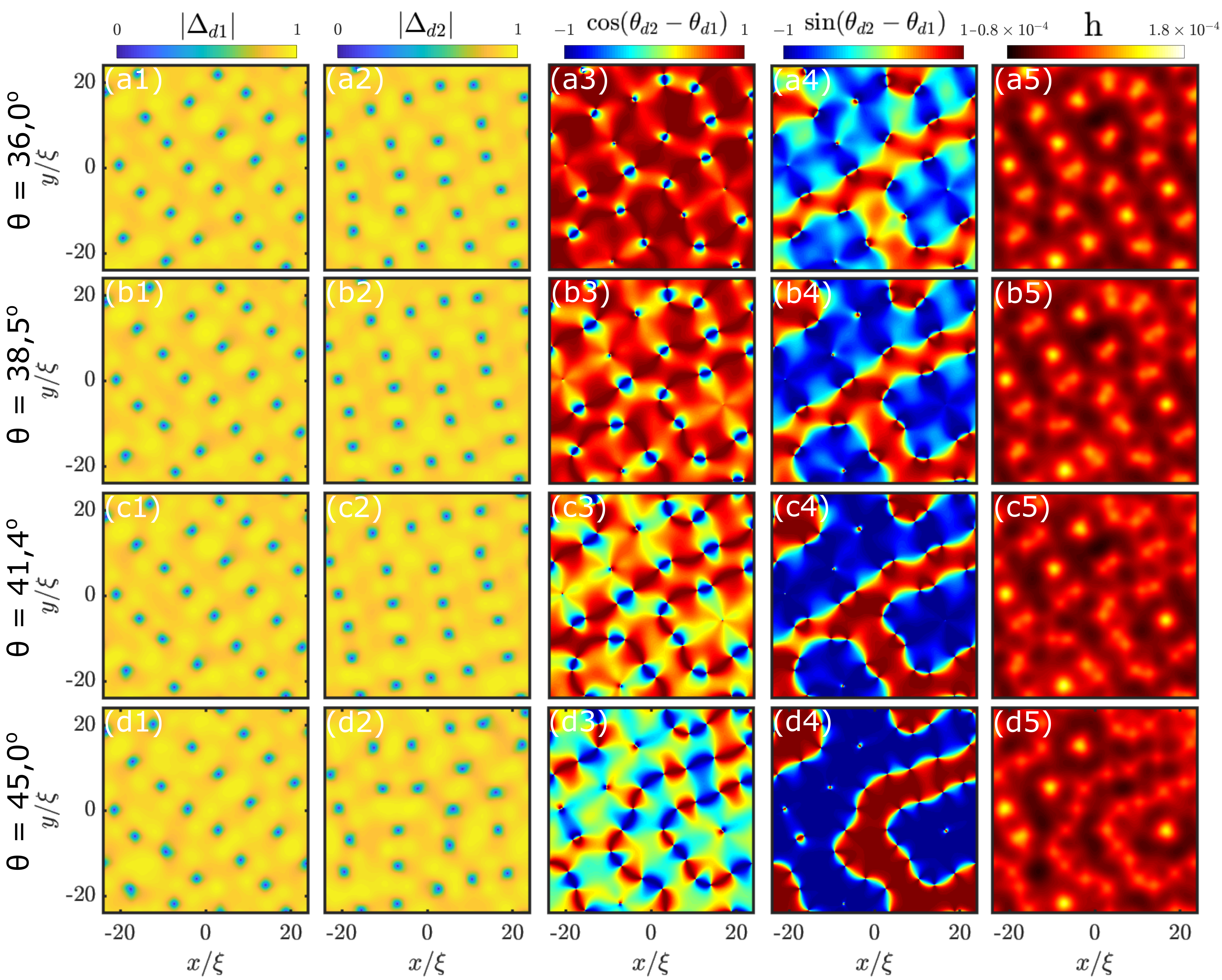}
    \caption{(Color online) Evolution of the vortex configuration when adiabatically increasing the twist angle from $\theta = 36^{\circ}$ to $45^{\circ}$. From left to right, the columns respectively show the magnitude of the $d$-wave component of the order parameter for unrotated and rotated layers, the cosine and sine of the phase difference between the order parameters in two layers, and the magnetic field distribution across the system.}
    \label{fig:fig7}
\end{figure*}

\begin{figure}[!t]
    \centering
    \includegraphics[width=\columnwidth]{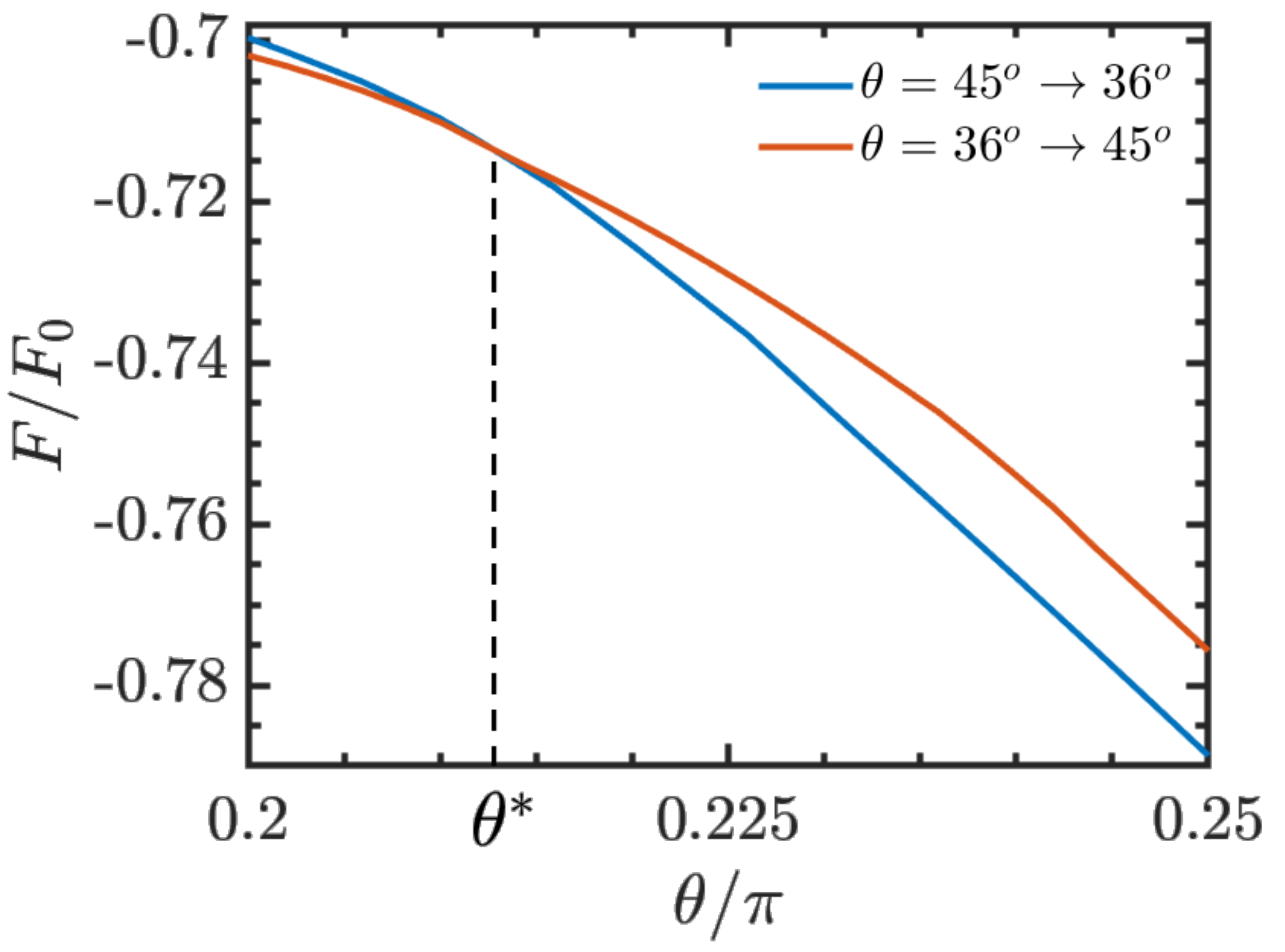}
    \caption{(Color online) Energy as a function of the twist angle $\theta$. The blue and red lines show the energy curves for the twist angle being decreased from $\theta = 45^{\circ}$ to $36^{\circ}$ and increased from $36^{\circ}$ to $45^{\circ}$, respectively.}
    \label{fig:fig8}
\end{figure}

\subsection{Vortex matter at the onset of the topological phase}
We start the description of the vortex matter for an angle close to $\theta_i$ (equivalent results are obtained for angles close to $\theta_f$). In Fig.~\ref{fig:fig3} we show vortex configurations found for $\theta = 36^{\circ}$, each row displaying minimum energy solutions for a different size of the unit cell. As discussed previously, the vortex cores in two layers are displaced from one another inside the topological phase, but each vortex of a given layer remains connected to its counterpart in the other layer. This is visible in the phase difference profile, suggesting existence of a phase soliton between the two vortex cores. Hereafter, we refer to this pair of connected vortices as the interlayer vortex pair. As can be seen in Fig.~\ref{fig:fig3}, inside the interlayer vortex pair we find phase difference $\theta_{d2}-\theta_{d1} = \pi$ between the condensates in two layers, and $\theta_{d2}-\theta_{d1} = 0$ outside of the pair.

As one object, the interlayer vortex pair displays skyrmionic properties, which can be described by first defining the pseudo-spin \cite{babaev2002}
\begin{equation}
    \bm{n} = \frac{\Delta_d^{\dagger}\bm{\sigma}\Delta_d}{\Delta_d^{\dagger}\Delta_d},
    \label{eq:eq8}
\end{equation}
with $\Delta_d = (\Delta_{(d1)},\Delta_{(d1)})$ and $\bm{\sigma} = (\sigma_1,\sigma_2,\sigma_3)$, where $\sigma_i$ is the Pauli matrix. With such pseudo-spin profile, one then calculates the topological charge of the system, defined as
\begin{equation}
    Q = \frac{1}{4\pi}\int\bm{n}\cdot\left(\frac{\partial\bm{n}}{\partial x}\times\frac{\partial\bm{n}}{\partial y}\right)dxdy .
    \label{eq:eq9}
\end{equation}

If we calculate the total topological charge for the three different configurations shown in Fig.~\ref{fig:fig3}, we obtain $Q = 24$ for each of them. As $24$ is also the number of flux quanta threading the shown unit cells, we conclude that each interlayer vortex pair is actually a skyrmionic object with a topological charge equal to $1$ (for a truly composite vortex, $Q=0$). We confirmed this further by calculating the topological charge not over the entire unit cell, but only around isolated interlayer pairs.

Within the dashed rectangles shown in Fig.~\ref{fig:fig3} we highlight vortex configurations characteristic of the topological phase for twist angles close to $\theta_i$. As can be seen from the cosine and sine of the phase difference between the condensate of each layer, two adjacent interlayer vortex pairs typically organize themselves into a larger correlated object. Inside the dashed rectangles in panels $(a1)-(a2)$ of Fig.~\ref{fig:fig3}, we can see that in one of the interlayer vortex pair (the one near the top of the rectangle) the vortex of the unrotated layer is on the right and the vortex of the rotated layer on the left. In the interlayer pair near the bottom of the rectangle, the vortex positions are interchanged. In other words, adjacent pairs of vortices in two layers are twisted with respect to each other. Same behavior can be easily verified in panels $(b1)-(b2)$ and $(c1)-(c2)$. After such organization of interlayer vortex pairs, their interlayer phase solitons become visibly connected, as seen in the dashed rectangle of the cosine of the phase difference in Fig.~\ref{fig:fig3}. Due to proximity and strong overlap between different phase domains, a supercurrent emerges surrounding the two interlayer vortex pairs, yields rather weak magnetic field. As a consequence, while the field profile of each pair is strong, and spatial correlation between them is rather obvious, the consequence of the phase connection between them is difficult to spot in the last column of Fig.~\ref{fig:fig3}.

\subsection{Vortex matter deep inside the topological phase} Finally we reveal the evolution of the vortex matter when the twisted bilayer is deeply inside the topological phase, i.e. for  twist angle $\theta \approx 45^{\circ}$ in the present case. As previously shown, for this $\theta$ the phase difference between the $d$-wave components of the order parameters in two layers is $\pi/2$ and we have a $d+id'$ superconducting state. Despite this particularity, the encountered characteristics of the vortex matter in this case can be related to the ones exhibited for other twist angles within the topological phase in the vicinity of $\theta=45^{\circ}$.

Fig.~\ref{fig:fig4} shows the order parameters of both layers, the cosine and the sine of the phase difference between the layers and the magnetic field profile for $\theta = 45^{\circ}$.
In this case, formation of interlayer vortex pairs with topological charge $Q = 1$ is still favorable, as highlighted by dashed rectangles in panels $(c1)-(c5)$ of Fig.~\ref{fig:fig4}. However, as highlighted by dashed rectangles in panels $(a1)-(a5)$, the organization of interlayer vortex pairs into larger objects is preferable. As a consequence, a new vortex configuration emerges - the skyrmionic chain. In this uniquely distinct state, instead of the small interlayer phase domains within individual interlayer vortex pairs, much larger domains are formed. Interlayer vortex pairs are interconnected along the domain wall, with a vortex core from one layer being in between two vortex cores of the other one, forming a closed chain of interlayer vortex pairs. Dashed rectangles in panels $(a1)-(a5)$ of Fig.~\ref{fig:fig4} exemplify one such structure, containing 11 interlayer vortex pairs in a single closed chain. In panel $(a5)$, one can see that such vortex chain is a very laterally extended object (nearly $40\xi\times40\xi$ in this case), and leaves a very clear and rather unusual signature in the magnetic field profile of the system. Here, the peaks of the magnetic field along the vortex chain are smaller in comparison with the isolated interlayer vortex pairs due to the fact that the distance between the vortex cores in two layers is significantly larger in the former case.

Once again, and as can be seen from the third and fourth columns in Fig.~\ref{fig:fig4}, the closed vortex chain separates two regions with different phase differences between the layers. Inside the vortex chain, the phase difference between the condensates is $\theta_{d2}-\theta_{d1} = -\pi/2$, while outside the chain $\theta_{d2}-\theta_{d1} = \pi/2$. The opposite is also possible: panels $(b1)-(b5)$ show such an example, where  $\theta_{d2}-\theta_{d1} = \pi/2$ inside the chain, while $\theta_{d2}-\theta_{d1} = -\pi/2$ outside. We note the difference from the case of the individual interlayer vortex pairs, harboring phase difference $\pi$ within them, with zero phase difference away from the pair.

If we now calculate the total topological charge around the vortex chains seen in panels $(a1)-(a5)$ and $(b1)-(b5)$, we obtain $Q = 11$ and $Q = 3$, respectively, reflecting the number of interlayer vortex pairs interconnected in the chain. These large values for the topological charge of such a novel object contrast the exclusively $Q = 1$ of the individual interlayer vortex pairs found for $\theta = 36^{\circ}$. This very different behavior for different twist angles emerges from the fact that, as discussed before, the domain wall separates regions with a phase difference equal to $0$ and $\pi$ for $\theta = 36^{\circ}$ and $-\pi/2$ and $\pi/2$ for $\theta = 45^{\circ}$.

To understand how the value of the phase difference inside the domains affects the topological charge, Fig.~\ref{fig:fig5} shows the energy of the homogeneous system as a function of $\theta_{d2}-\theta_{d1}$ for different values of the twist angle $\theta$. At the onset of the topological phase (blue curve in Fig.~\ref{fig:fig5}), the energy of the system is largest when the phase difference is $\pi$. Therefore, larger splitting within each formed interlayer vortex pair costs energy, and their interconnection into larger objects is not energetically favorable. Notice that, as discussed in Fig.~\ref{fig:fig1}, the phase difference equal to $0$ does not yield the free energy minimum in the homogeneous state of the system for $\theta$ at which the topological state is stable. In the presence of magnetic field, the formation of interlayer vortex pairs re-stabilizes the zero phase difference in a large part of the superconductor for $\theta$ close to $\theta_i$ and $\theta_f$.

On the other hand, deep in the topological phase (for $\theta = 45^{\circ}$), a degenerate lowest energy homogeneous state is found for phase difference equal to either $-\pi/2$ or $\pi/2$, explaining the tendency to formation of coexisting domains with such phase differences. The resulting long domain walls would cost energy, but not in the presence of magnetic field when they are decorated by the skyrmionic vortex chains. 

For $\theta$ values in the vicinity of $45^{\circ}$, represented in Fig.~\ref{fig:fig5} by $\theta = 42^{\circ}$, the free-energy minima no longer occur at $-\pi/2$ and $\pi/2$ but shift to lower phase differences and become shallower (cf. Fig.~\ref{fig:fig5}). Nevertheless, the system still presents the vortex chains dividing the superconductor in regions with phase differences $-\pi/2$ and $\pi/2$, since vortices require a total phase difference $\pi$ across the domain wall on which they reside. Once both values of the phase difference possess the same free energy, the long domain walls described above are also present, with the same size as the ones for $\theta = 45^{\circ}$.

\subsection{Transitions between the topological vortex matter with the interlayer twist}

Complementary, it seems relevant to discuss in which manner the above-described characteristic skyrmionic states in the topological phase evolve as one continuously varies the twist angle. To capture this behavior, we follow two distinct procedures. In the first, we start deep in the topological phase, i.e. at a twist angle $\theta = 45^{\circ}$ and a skyrmionic vortex chain as the initial state of the simulation. We then ``adiabatically'' decrease the twist angle down to $36^{\circ}$, in decrements of $0.1^{\circ}$, recording the evolution of the stable solution (which is no longer necessarily the lowest-energy state). In Fig.~\ref{fig:fig6} we show the selected vortex configurations obtained during this procedure. Starting from the skyrmionic vortex chain (panels $(a1)-(a5)$), we see that when the twist angle is decreased to $\theta = 42,1^{\circ}$ (panels $(b1)-(b5)$), the contour of the chain can still be seen in the magnetic field profile of the system. At the same time, the sine of the phase difference shows that the vortex chain still splits the superconducting landscape in regions with interlayer phase difference equal to either $\pi/2$ and $-\pi/2$. However, vortices in each layer start to group in pairs, as reflected in double peaks appearing in the magnetic field profile along the chain. This indicates the onset of the transition from the skyrmionic vortex chain to the skyrmionic state with separate interlayer vortex pairs. Such a transition is more apparent for $\theta = 38,9^{\circ}$ (panels $(c1)-(c5)$). Although the magnetic field contour of the vortex chain can still be visualized in this case, the separation of the superconductor into regions with different interlayer phase differences becomes less clear. Finally, at $\theta = 36^{\circ}$ (panels $(d1)-(d5)$) the transition between the two skyrmionic states is completed and the system displays an arrangement of dissociated individual interlayer vortex pairs, surrounded by a landscape of near-zero interlayer phase difference.

Along the opposite route, we start with the lowest-energy state with interlayer vortex pairs for $\theta = 36^{\circ}$ (shown in panels $(a1)-(a5)$ of Fig.~\ref{fig:fig7}) as the initial state of our system and then gradually increase the twist angle up to $\theta = 45^{\circ}$. As can be conveniently seen from the third and fourth columns of Fig.~\ref{fig:fig7}, the state gradually changes from the topological phase with regions of interlayer phase difference either $0$ or $\pi$ to another one with regions of interlayer phase differences either $\pi/2$ or $-\pi/2$. As the twist angle is increased, the vortices forming an interlayer vortex pair slowly separate from each other. One sees this by comparing the magnetic profile in panels $(b5)$ and $(c5)$, where the double peak characteristic of a vortex pair becomes smeared. For twist angles in the vicinity of $\theta = 45^{\circ}$, this culminates in the formation of the skyrmionic vortex chain state, as we display in panels $(d1)-(d5)$.

Both discussed transitions between the two different topological skyrmionic vortex states occur through the second-order relocation of vortex cores in each layer. This is further evidenced in Fig.~\ref{fig:fig8}, where we show the energy of the system as a function of the twist angle for the cases where $\theta$ is decreased from $45^{\circ}$ to $36^{\circ}$ (blue line, cf. Fig.~\ref{fig:fig6}), and increased from $36^{\circ}$ to $45^{\circ}$ (red line, cf. Fig.~\ref{fig:fig7}). Notably, the two detected characteristic skyrmionic vortex states in the topological phase of the system cross in energy at a twist angle $\theta^* \approx 38^{\circ}$. To emphasize again, the interlayer vortex pairs are energetically favorable for interlayer twist below this angle, whereas the states containing skyrmionic vortex chains become favorable for $\theta>\theta^*$. Obviously the exact value of $\theta^*$ will depend on the details of the simulation (size of the unit cell, magnetic field), but we can safely generalize this result to conclude that skyrmionic vortex chains should be observable in the larger portion of the twist range where topological phase is expected. 

\section{\label{sec:sec4}Conclusion}

To summarize, we have analyzed the vortex configurations emerging in a twisted bilayer composed of superconducting monolayers with $d$-wave pairing - motivated by prospects of such realizations using e.g. Bi$_2$Sr$_2$CaCu$_2$O$_{8+\delta}$. In such a system, the phase difference between the superconducting order parameters in two layers depends on the twist angle $\theta$, with a topological state with a non-trivial phase difference emerging for a range of angles around $\theta=45^\circ$. In that topological phase, the superconducting state exhibits broken time-reversal symmetry, giving rise to skyrmionic vortex configurations with topological charge not equal to zero. We revealed and characterized those nontrivial vortex states, and discussed their detectable differences when compared to usual vortex lattice found for twist angles outside the topological range. Based on those clearly discernible differences, we argue that direct experimental observation of skyrmionic vortex states can be used as a smoking gun to detect topological superconductivity in such systems.

In addition, we showed that the skyrmionic vortex matter also evolves with the twist angle inside the topological phase. Namely, we have identified two distinct types of skyrmionic states. At the onset of the topological state, the system prefers to preserve same phase of the order parameter in two layers, so the broken reversal symmetry reflects solely in formation of the interlayer vortex pairs. Each of this pairs carries a unit of topological charge, and hosts phase difference $\pi$ between the coupled superconducting layers. As the twist angle is varied towards $45^\circ$ and one is deeper in the topological state, the phase difference of $\pm\pi/2$ becomes energetically favorable. As a consequence, the interlayer vortex pairs interconnect into extended closed chains, separating the regions of the sample with phase difference either $-\pi/2$ or $\pi/2$. Such chains can easily exhibit lateral extent on the micron scale, and carry topological charge equal to the number of vortices interconnected in the chain. Once again, we emphasize that each of the two types of skyrmionic flux objects leaves a clear signature in the spatial profile of the magnetic field across the system, but will also host uniquely related local density of states and bound states detectable by e.g. Scanning Tunneling Microscopy. The calculation of such states is left as a prospect for further work, being beyond the capability of the present Ginzburg-Landau analysis (Bogolyubov-deGennes approach is a viable alternative \cite{zhang2016,ying2018}.
Another interesting outlook is to adapt the here-presented Ginzburg-Landau formalism to the cases of other pairing symmetries that may arise in the twisted bilayers of present interest, so to classify the emergent vortex matter according to the symmetries at hand - and thereby enable conclusive identification of the pairing symmetry in experimental systems through visualization of the vortex states - complementary to other existing efforts (see e.g. \cite{fengcheng2019}).

\begin{acknowledgments}
This work has been supported by the Research Foundation-Flanders (FWO-Vlaanderen), Special Research Funds of the University of Antwerp (BOF-UA), and Brazilian Agency FAPESP (grant numbers 20/03947-2 and 20/10058-0).
\end{acknowledgments}

\bibliography{bibliography}

\begin{thebibliography}{31}%
\makeatletter
\providecommand \@ifxundefined [1]{%
 \@ifx{#1\undefined}
}%
\providecommand \@ifnum [1]{%
 \ifnum #1\expandafter \@firstoftwo
 \else \expandafter \@secondoftwo
 \fi
}%
\providecommand \@ifx [1]{%
 \ifx #1\expandafter \@firstoftwo
 \else \expandafter \@secondoftwo
 \fi
}%
\providecommand \natexlab [1]{#1}%
\providecommand \enquote  [1]{``#1''}%
\providecommand \bibnamefont  [1]{#1}%
\providecommand \bibfnamefont [1]{#1}%
\providecommand \citenamefont [1]{#1}%
\providecommand \href@noop [0]{\@secondoftwo}%
\providecommand \href [0]{\begingroup \@sanitize@url \@href}%
\providecommand \@href[1]{\@@startlink{#1}\@@href}%
\providecommand \@@href[1]{\endgroup#1\@@endlink}%
\providecommand \@sanitize@url [0]{\catcode `\\12\catcode `\$12\catcode
  `\&12\catcode `\#12\catcode `\^12\catcode `\_12\catcode `\%12\relax}%
\providecommand \@@startlink[1]{}%
\providecommand \@@endlink[0]{}%
\providecommand \url  [0]{\begingroup\@sanitize@url \@url }%
\providecommand \@url [1]{\endgroup\@href {#1}{\urlprefix }}%
\providecommand \urlprefix  [0]{URL }%
\providecommand \Eprint [0]{\href }%
\providecommand \doibase [0]{https://doi.org/}%
\providecommand \selectlanguage [0]{\@gobble}%
\providecommand \bibinfo  [0]{\@secondoftwo}%
\providecommand \bibfield  [0]{\@secondoftwo}%
\providecommand \translation [1]{[#1]}%
\providecommand \BibitemOpen [0]{}%
\providecommand \bibitemStop [0]{}%
\providecommand \bibitemNoStop [0]{.\EOS\space}%
\providecommand \EOS [0]{\spacefactor3000\relax}%
\providecommand \BibitemShut  [1]{\csname bibitem#1\endcsname}%
\let\auto@bib@innerbib\@empty
\bibitem [{\citenamefont {Kallin}\ and\ \citenamefont
  {Berlinsky}(2016)}]{kallin2016}%
  \BibitemOpen
  \bibfield  {author} {\bibinfo {author} {\bibfnamefont {C.}~\bibnamefont
  {Kallin}}\ and\ \bibinfo {author} {\bibfnamefont {J.}~\bibnamefont
  {Berlinsky}},\ }\bibfield  {title} {\bibinfo {title} {Chiral
  superconductors},\ }\href@noop {} {\bibfield  {journal} {\bibinfo  {journal}
  {Reports on Progress in Physics}\ }\textbf {\bibinfo {volume} {79}},\
  \bibinfo {pages} {054502} (\bibinfo {year} {2016})}\BibitemShut {NoStop}%
\bibitem [{\citenamefont {Sigrist}\ and\ \citenamefont
  {Ueda}(1991)}]{sigrist1991}%
  \BibitemOpen
  \bibfield  {author} {\bibinfo {author} {\bibfnamefont {M.}~\bibnamefont
  {Sigrist}}\ and\ \bibinfo {author} {\bibfnamefont {K.}~\bibnamefont {Ueda}},\
  }\bibfield  {title} {\bibinfo {title} {Phenomenological theory of
  unconventional superconductivity},\ }\href@noop {} {\bibfield  {journal}
  {\bibinfo  {journal} {Reviews of Modern physics}\ }\textbf {\bibinfo {volume}
  {63}},\ \bibinfo {pages} {239} (\bibinfo {year} {1991})}\BibitemShut
  {NoStop}%
\bibitem [{\citenamefont {Vojta}\ \emph {et~al.}(2000)\citenamefont {Vojta},
  \citenamefont {Zhang},\ and\ \citenamefont {Sachdev}}]{vojta2000}%
  \BibitemOpen
  \bibfield  {author} {\bibinfo {author} {\bibfnamefont {M.}~\bibnamefont
  {Vojta}}, \bibinfo {author} {\bibfnamefont {Y.}~\bibnamefont {Zhang}},\ and\
  \bibinfo {author} {\bibfnamefont {S.}~\bibnamefont {Sachdev}},\ }\bibfield
  {title} {\bibinfo {title} {Quantum phase transitions in d-wave
  superconductors},\ }\href@noop {} {\bibfield  {journal} {\bibinfo  {journal}
  {Physical review letters}\ }\textbf {\bibinfo {volume} {85}},\ \bibinfo
  {pages} {4940} (\bibinfo {year} {2000})}\BibitemShut {NoStop}%
\bibitem [{\citenamefont {Stone}\ and\ \citenamefont {Roy}(2004)}]{stone2004}%
  \BibitemOpen
  \bibfield  {author} {\bibinfo {author} {\bibfnamefont {M.}~\bibnamefont
  {Stone}}\ and\ \bibinfo {author} {\bibfnamefont {R.}~\bibnamefont {Roy}},\
  }\bibfield  {title} {\bibinfo {title} {Edge modes, edge currents, and gauge
  invariance in p x+ i p y superfluids and superconductors},\ }\href@noop {}
  {\bibfield  {journal} {\bibinfo  {journal} {Physical Review B}\ }\textbf
  {\bibinfo {volume} {69}},\ \bibinfo {pages} {184511} (\bibinfo {year}
  {2004})}\BibitemShut {NoStop}%
\bibitem [{\citenamefont {Read}\ and\ \citenamefont {Green}(2000)}]{read2000}%
  \BibitemOpen
  \bibfield  {author} {\bibinfo {author} {\bibfnamefont {N.}~\bibnamefont
  {Read}}\ and\ \bibinfo {author} {\bibfnamefont {D.}~\bibnamefont {Green}},\
  }\bibfield  {title} {\bibinfo {title} {Paired states of fermions in two
  dimensions with breaking of parity and time-reversal symmetries and the
  fractional quantum hall effect},\ }\href@noop {} {\bibfield  {journal}
  {\bibinfo  {journal} {Physical Review B}\ }\textbf {\bibinfo {volume} {61}},\
  \bibinfo {pages} {10267} (\bibinfo {year} {2000})}\BibitemShut {NoStop}%
\bibitem [{\citenamefont {Garaud}\ and\ \citenamefont
  {Babaev}(2012)}]{garaud2012}%
  \BibitemOpen
  \bibfield  {author} {\bibinfo {author} {\bibfnamefont {J.}~\bibnamefont
  {Garaud}}\ and\ \bibinfo {author} {\bibfnamefont {E.}~\bibnamefont
  {Babaev}},\ }\bibfield  {title} {\bibinfo {title} {Skyrmionic state and
  stable half-quantum vortices in chiral p-wave superconductors},\ }\href@noop
  {} {\bibfield  {journal} {\bibinfo  {journal} {Physical Review B}\ }\textbf
  {\bibinfo {volume} {86}},\ \bibinfo {pages} {060514} (\bibinfo {year}
  {2012})}\BibitemShut {NoStop}%
\bibitem [{\citenamefont {Jang}\ \emph {et~al.}(2011)\citenamefont {Jang},
  \citenamefont {Ferguson}, \citenamefont {Vakaryuk}, \citenamefont {Budakian},
  \citenamefont {Chung}, \citenamefont {Goldbart},\ and\ \citenamefont
  {Maeno}}]{jang2011}%
  \BibitemOpen
  \bibfield  {author} {\bibinfo {author} {\bibfnamefont {J.}~\bibnamefont
  {Jang}}, \bibinfo {author} {\bibfnamefont {D.}~\bibnamefont {Ferguson}},
  \bibinfo {author} {\bibfnamefont {V.}~\bibnamefont {Vakaryuk}}, \bibinfo
  {author} {\bibfnamefont {R.}~\bibnamefont {Budakian}}, \bibinfo {author}
  {\bibfnamefont {S.}~\bibnamefont {Chung}}, \bibinfo {author} {\bibfnamefont
  {P.}~\bibnamefont {Goldbart}},\ and\ \bibinfo {author} {\bibfnamefont
  {Y.}~\bibnamefont {Maeno}},\ }\bibfield  {title} {\bibinfo {title}
  {Observation of half-height magnetization steps in sr2ruo4},\ }\href@noop {}
  {\bibfield  {journal} {\bibinfo  {journal} {Science}\ }\textbf {\bibinfo
  {volume} {331}},\ \bibinfo {pages} {186} (\bibinfo {year}
  {2011})}\BibitemShut {NoStop}%
\bibitem [{\citenamefont {Zyuzin}\ \emph {et~al.}(2017)\citenamefont {Zyuzin},
  \citenamefont {Garaud},\ and\ \citenamefont {Babaev}}]{zyuzin2017}%
  \BibitemOpen
  \bibfield  {author} {\bibinfo {author} {\bibfnamefont {A.}~\bibnamefont
  {Zyuzin}}, \bibinfo {author} {\bibfnamefont {J.}~\bibnamefont {Garaud}},\
  and\ \bibinfo {author} {\bibfnamefont {E.}~\bibnamefont {Babaev}},\
  }\bibfield  {title} {\bibinfo {title} {Nematic skyrmions in odd-parity
  superconductors},\ }\href@noop {} {\bibfield  {journal} {\bibinfo  {journal}
  {Physical review letters}\ }\textbf {\bibinfo {volume} {119}},\ \bibinfo
  {pages} {167001} (\bibinfo {year} {2017})}\BibitemShut {NoStop}%
\bibitem [{\citenamefont {Becerra}\ \emph {et~al.}(2016)\citenamefont
  {Becerra}, \citenamefont {Sardella}, \citenamefont {Peeters},\ and\
  \citenamefont {Milo{\v{s}}evi{\'c}}}]{becerra2016}%
  \BibitemOpen
  \bibfield  {author} {\bibinfo {author} {\bibfnamefont {V.~F.}\ \bibnamefont
  {Becerra}}, \bibinfo {author} {\bibfnamefont {E.}~\bibnamefont {Sardella}},
  \bibinfo {author} {\bibfnamefont {F.}~\bibnamefont {Peeters}},\ and\ \bibinfo
  {author} {\bibfnamefont {M.}~\bibnamefont {Milo{\v{s}}evi{\'c}}},\ }\bibfield
   {title} {\bibinfo {title} {Vortical versus skyrmionic states in mesoscopic
  p-wave superconductors},\ }\href@noop {} {\bibfield  {journal} {\bibinfo
  {journal} {Physical Review B}\ }\textbf {\bibinfo {volume} {93}},\ \bibinfo
  {pages} {014518} (\bibinfo {year} {2016})}\BibitemShut {NoStop}%
\bibitem [{\citenamefont {Milo{\v{s}}evi{\'{c}}}\ and\ \citenamefont
  {Perali}(2015)}]{milosevic2015}%
  \BibitemOpen
  \bibfield  {author} {\bibinfo {author} {\bibfnamefont {M.~V.}\ \bibnamefont
  {Milo{\v{s}}evi{\'{c}}}}\ and\ \bibinfo {author} {\bibfnamefont
  {A.}~\bibnamefont {Perali}},\ }\bibfield  {title} {\bibinfo {title} {Emergent
  phenomena in multicomponent superconductivity: an introduction to the focus
  issue},\ }\href {https://doi.org/10.1088/0953-2048/28/6/060201} {\bibfield
  {journal} {\bibinfo  {journal} {Superconductor Science and Technology}\
  }\textbf {\bibinfo {volume} {28}},\ \bibinfo {pages} {060201} (\bibinfo
  {year} {2015})}\BibitemShut {NoStop}%
\bibitem [{\citenamefont {Tanaka}(2015)}]{tanaka2015}%
  \BibitemOpen
  \bibfield  {author} {\bibinfo {author} {\bibfnamefont {Y.}~\bibnamefont
  {Tanaka}},\ }\bibfield  {title} {\bibinfo {title} {Multicomponent
  superconductivity based on multiband superconductors},\ }\href@noop {}
  {\bibfield  {journal} {\bibinfo  {journal} {Superconductor Science and
  Technology}\ }\textbf {\bibinfo {volume} {28}},\ \bibinfo {pages} {034002}
  (\bibinfo {year} {2015})}\BibitemShut {NoStop}%
\bibitem [{\citenamefont {Babaev}\ and\ \citenamefont
  {Speight}(2005)}]{babaev2005}%
  \BibitemOpen
  \bibfield  {author} {\bibinfo {author} {\bibfnamefont {E.}~\bibnamefont
  {Babaev}}\ and\ \bibinfo {author} {\bibfnamefont {M.}~\bibnamefont
  {Speight}},\ }\bibfield  {title} {\bibinfo {title} {Semi-meissner state and
  neither type-i nor type-ii superconductivity in multicomponent
  superconductors},\ }\href@noop {} {\bibfield  {journal} {\bibinfo  {journal}
  {Physical Review B}\ }\textbf {\bibinfo {volume} {72}},\ \bibinfo {pages}
  {180502} (\bibinfo {year} {2005})}\BibitemShut {NoStop}%
\bibitem [{\citenamefont {Lin}\ and\ \citenamefont {Hu}(2011)}]{lin2011}%
  \BibitemOpen
  \bibfield  {author} {\bibinfo {author} {\bibfnamefont {S.-Z.}\ \bibnamefont
  {Lin}}\ and\ \bibinfo {author} {\bibfnamefont {X.}~\bibnamefont {Hu}},\
  }\bibfield  {title} {\bibinfo {title} {Vortex states and the phase diagram of
  a multiple-component ginzburg-landau theory with competing repulsive and
  attractive vortex interactions},\ }\href@noop {} {\bibfield  {journal}
  {\bibinfo  {journal} {Physical Review B}\ }\textbf {\bibinfo {volume} {84}},\
  \bibinfo {pages} {214505} (\bibinfo {year} {2011})}\BibitemShut {NoStop}%
\bibitem [{\citenamefont {Tanaka}(2001)}]{tanaka2001}%
  \BibitemOpen
  \bibfield  {author} {\bibinfo {author} {\bibfnamefont {Y.}~\bibnamefont
  {Tanaka}},\ }\bibfield  {title} {\bibinfo {title} {Soliton in two-band
  superconductor},\ }\href@noop {} {\bibfield  {journal} {\bibinfo  {journal}
  {Physical review letters}\ }\textbf {\bibinfo {volume} {88}},\ \bibinfo
  {pages} {017002} (\bibinfo {year} {2001})}\BibitemShut {NoStop}%
\bibitem [{\citenamefont {Sato}\ and\ \citenamefont {Ando}(2017)}]{sato2017}%
  \BibitemOpen
  \bibfield  {author} {\bibinfo {author} {\bibfnamefont {M.}~\bibnamefont
  {Sato}}\ and\ \bibinfo {author} {\bibfnamefont {Y.}~\bibnamefont {Ando}},\
  }\bibfield  {title} {\bibinfo {title} {Topological superconductors: a
  review},\ }\href@noop {} {\bibfield  {journal} {\bibinfo  {journal} {Reports
  on Progress in Physics}\ }\textbf {\bibinfo {volume} {80}},\ \bibinfo {pages}
  {076501} (\bibinfo {year} {2017})}\BibitemShut {NoStop}%
\bibitem [{\citenamefont {Sarma}\ \emph {et~al.}(2015)\citenamefont {Sarma},
  \citenamefont {Freedman},\ and\ \citenamefont {Nayak}}]{sarma2015}%
  \BibitemOpen
  \bibfield  {author} {\bibinfo {author} {\bibfnamefont {S.~D.}\ \bibnamefont
  {Sarma}}, \bibinfo {author} {\bibfnamefont {M.}~\bibnamefont {Freedman}},\
  and\ \bibinfo {author} {\bibfnamefont {C.}~\bibnamefont {Nayak}},\ }\bibfield
   {title} {\bibinfo {title} {Majorana zero modes and topological quantum
  computation},\ }\href@noop {} {\bibfield  {journal} {\bibinfo  {journal} {npj
  Quantum Information}\ }\textbf {\bibinfo {volume} {1}},\ \bibinfo {pages} {1}
  (\bibinfo {year} {2015})}\BibitemShut {NoStop}%
\bibitem [{\citenamefont {Volovik}(1997)}]{volovik1997}%
  \BibitemOpen
  \bibfield  {author} {\bibinfo {author} {\bibfnamefont {G.}~\bibnamefont
  {Volovik}},\ }\bibfield  {title} {\bibinfo {title} {On edge states in
  superconductors with time inversion symmetry breaking},\ }\href@noop {}
  {\bibfield  {journal} {\bibinfo  {journal} {Journal of Experimental and
  Theoretical Physics Letters}\ }\textbf {\bibinfo {volume} {66}},\ \bibinfo
  {pages} {522} (\bibinfo {year} {1997})}\BibitemShut {NoStop}%
\bibitem [{\citenamefont {Volovik}(1999)}]{volovik1999}%
  \BibitemOpen
  \bibfield  {author} {\bibinfo {author} {\bibfnamefont {G.}~\bibnamefont
  {Volovik}},\ }\bibfield  {title} {\bibinfo {title} {Fermion zero modes on
  vortices in chiral superconductors},\ }\href@noop {} {\bibfield  {journal}
  {\bibinfo  {journal} {Journal of Experimental and Theoretical Physics
  Letters}\ }\textbf {\bibinfo {volume} {70}},\ \bibinfo {pages} {609}
  (\bibinfo {year} {1999})}\BibitemShut {NoStop}%
\bibitem [{\citenamefont {Ivanov}(2001)}]{ivanov2001}%
  \BibitemOpen
  \bibfield  {author} {\bibinfo {author} {\bibfnamefont {D.~A.}\ \bibnamefont
  {Ivanov}},\ }\bibfield  {title} {\bibinfo {title} {Non-abelian statistics of
  half-quantum vortices in p-wave superconductors},\ }\href@noop {} {\bibfield
  {journal} {\bibinfo  {journal} {Physical review letters}\ }\textbf {\bibinfo
  {volume} {86}},\ \bibinfo {pages} {268} (\bibinfo {year} {2001})}\BibitemShut
  {NoStop}%
\bibitem [{\citenamefont {Can}\ \emph {et~al.}(2021)\citenamefont {Can},
  \citenamefont {Tummuru}, \citenamefont {Day}, \citenamefont {Elfimov},
  \citenamefont {Damascelli},\ and\ \citenamefont {Franz}}]{can2021}%
  \BibitemOpen
  \bibfield  {author} {\bibinfo {author} {\bibfnamefont {O.}~\bibnamefont
  {Can}}, \bibinfo {author} {\bibfnamefont {T.}~\bibnamefont {Tummuru}},
  \bibinfo {author} {\bibfnamefont {R.~P.}\ \bibnamefont {Day}}, \bibinfo
  {author} {\bibfnamefont {I.}~\bibnamefont {Elfimov}}, \bibinfo {author}
  {\bibfnamefont {A.}~\bibnamefont {Damascelli}},\ and\ \bibinfo {author}
  {\bibfnamefont {M.}~\bibnamefont {Franz}},\ }\bibfield  {title} {\bibinfo
  {title} {High-temperature topological superconductivity in twisted
  double-layer copper oxides},\ }\href@noop {} {\bibfield  {journal} {\bibinfo
  {journal} {Nature Physics}\ }\textbf {\bibinfo {volume} {17}},\ \bibinfo
  {pages} {519} (\bibinfo {year} {2021})}\BibitemShut {NoStop}%
\bibitem [{\citenamefont {Yu}\ \emph {et~al.}(2019)\citenamefont {Yu},
  \citenamefont {Ma}, \citenamefont {Cai}, \citenamefont {Zhong}, \citenamefont
  {Ye}, \citenamefont {Shen}, \citenamefont {Gu}, \citenamefont {Chen},\ and\
  \citenamefont {Zhang}}]{yu2019}%
  \BibitemOpen
  \bibfield  {author} {\bibinfo {author} {\bibfnamefont {Y.}~\bibnamefont
  {Yu}}, \bibinfo {author} {\bibfnamefont {L.}~\bibnamefont {Ma}}, \bibinfo
  {author} {\bibfnamefont {P.}~\bibnamefont {Cai}}, \bibinfo {author}
  {\bibfnamefont {R.}~\bibnamefont {Zhong}}, \bibinfo {author} {\bibfnamefont
  {C.}~\bibnamefont {Ye}}, \bibinfo {author} {\bibfnamefont {J.}~\bibnamefont
  {Shen}}, \bibinfo {author} {\bibfnamefont {G.~D.}\ \bibnamefont {Gu}},
  \bibinfo {author} {\bibfnamefont {X.~H.}\ \bibnamefont {Chen}},\ and\
  \bibinfo {author} {\bibfnamefont {Y.}~\bibnamefont {Zhang}},\ }\bibfield
  {title} {\bibinfo {title} {High-temperature superconductivity in monolayer
  bi2sr2cacu2o8+ $\delta$},\ }\href@noop {} {\bibfield  {journal} {\bibinfo
  {journal} {Nature}\ }\textbf {\bibinfo {volume} {575}},\ \bibinfo {pages}
  {156} (\bibinfo {year} {2019})}\BibitemShut {NoStop}%
\bibitem [{\citenamefont {Babaev}\ \emph
  {et~al.}(2002{\natexlab{a}})\citenamefont {Babaev}, \citenamefont {Faddeev},\
  and\ \citenamefont {Niemi}}]{babaev2002b}%
  \BibitemOpen
  \bibfield  {author} {\bibinfo {author} {\bibfnamefont {E.}~\bibnamefont
  {Babaev}}, \bibinfo {author} {\bibfnamefont {L.~D.}\ \bibnamefont
  {Faddeev}},\ and\ \bibinfo {author} {\bibfnamefont {A.~J.}\ \bibnamefont
  {Niemi}},\ }\bibfield  {title} {\bibinfo {title} {Hidden symmetry and knot
  solitons in a charged two-condensate bose system},\ }\href@noop {} {\bibfield
   {journal} {\bibinfo  {journal} {Physical Review B}\ }\textbf {\bibinfo
  {volume} {65}},\ \bibinfo {pages} {100512} (\bibinfo {year}
  {2002}{\natexlab{a}})}\BibitemShut {NoStop}%
\bibitem [{\citenamefont {Garaud}\ \emph {et~al.}(2013)\citenamefont {Garaud},
  \citenamefont {Carlstr\"om}, \citenamefont {Babaev},\ and\ \citenamefont
  {Speight}}]{garaud2013}%
  \BibitemOpen
  \bibfield  {author} {\bibinfo {author} {\bibfnamefont {J.}~\bibnamefont
  {Garaud}}, \bibinfo {author} {\bibfnamefont {J.}~\bibnamefont {Carlstr\"om}},
  \bibinfo {author} {\bibfnamefont {E.}~\bibnamefont {Babaev}},\ and\ \bibinfo
  {author} {\bibfnamefont {M.}~\bibnamefont {Speight}},\ }\bibfield  {title}
  {\bibinfo {title} {Chiral $\mathbb{C}{P}^{2}$ skyrmions in three-band
  superconductors},\ }\href {https://doi.org/10.1103/PhysRevB.87.014507}
  {\bibfield  {journal} {\bibinfo  {journal} {Phys. Rev. B}\ }\textbf {\bibinfo
  {volume} {87}},\ \bibinfo {pages} {014507} (\bibinfo {year}
  {2013})}\BibitemShut {NoStop}%
\bibitem [{\citenamefont {Benfenati}\ \emph {et~al.}(2022)\citenamefont
  {Benfenati}, \citenamefont {Barkman},\ and\ \citenamefont
  {Babaev}}]{benfenati2022}%
  \BibitemOpen
  \bibfield  {author} {\bibinfo {author} {\bibfnamefont {A.}~\bibnamefont
  {Benfenati}}, \bibinfo {author} {\bibfnamefont {M.}~\bibnamefont {Barkman}},\
  and\ \bibinfo {author} {\bibfnamefont {E.}~\bibnamefont {Babaev}},\
  }\bibfield  {title} {\bibinfo {title} {Demonstration of $\mathbb{C}{P}^{2}$
  skyrmions in three-band superconductors by self-consistent solutions to a
  bogoliubov-de gennes model},\ }\href@noop {} {\bibfield  {journal} {\bibinfo
  {journal} {arXiv preprint arXiv:2204.05242}\ } (\bibinfo {year}
  {2022})}\BibitemShut {NoStop}%
\bibitem [{\citenamefont {Zhang}\ \emph {et~al.}(2020)\citenamefont {Zhang},
  \citenamefont {Zhang}, \citenamefont {Zha}, \citenamefont
  {Milo{\v{s}}evi{\'c}},\ and\ \citenamefont {Zhou}}]{zhang2020}%
  \BibitemOpen
  \bibfield  {author} {\bibinfo {author} {\bibfnamefont {L.-F.}\ \bibnamefont
  {Zhang}}, \bibinfo {author} {\bibfnamefont {Y.-Y.}\ \bibnamefont {Zhang}},
  \bibinfo {author} {\bibfnamefont {G.-Q.}\ \bibnamefont {Zha}}, \bibinfo
  {author} {\bibfnamefont {M.}~\bibnamefont {Milo{\v{s}}evi{\'c}}},\ and\
  \bibinfo {author} {\bibfnamefont {S.-P.}\ \bibnamefont {Zhou}},\ }\bibfield
  {title} {\bibinfo {title} {Skyrmionic chains and lattices in s+ i d
  superconductors},\ }\href@noop {} {\bibfield  {journal} {\bibinfo  {journal}
  {Physical Review B}\ }\textbf {\bibinfo {volume} {101}},\ \bibinfo {pages}
  {064501} (\bibinfo {year} {2020})}\BibitemShut {NoStop}%
\bibitem [{\citenamefont {Ren}\ \emph {et~al.}(1995)\citenamefont {Ren},
  \citenamefont {Xu},\ and\ \citenamefont {Ting}}]{ren1995}%
  \BibitemOpen
  \bibfield  {author} {\bibinfo {author} {\bibfnamefont {Y.}~\bibnamefont
  {Ren}}, \bibinfo {author} {\bibfnamefont {J.-H.}\ \bibnamefont {Xu}},\ and\
  \bibinfo {author} {\bibfnamefont {C.}~\bibnamefont {Ting}},\ }\bibfield
  {title} {\bibinfo {title} {Ginzburg-landau equations and vortex structure of
  a d x 2- y 2 superconductor},\ }\href@noop {} {\bibfield  {journal} {\bibinfo
   {journal} {Physical review letters}\ }\textbf {\bibinfo {volume} {74}},\
  \bibinfo {pages} {3680} (\bibinfo {year} {1995})}\BibitemShut {NoStop}%
\bibitem [{\citenamefont {Doria}\ \emph {et~al.}(1989)\citenamefont {Doria},
  \citenamefont {Gubernatis},\ and\ \citenamefont {Rainer}}]{doria1989}%
  \BibitemOpen
  \bibfield  {author} {\bibinfo {author} {\bibfnamefont {M.~M.}\ \bibnamefont
  {Doria}}, \bibinfo {author} {\bibfnamefont {J.}~\bibnamefont {Gubernatis}},\
  and\ \bibinfo {author} {\bibfnamefont {D.}~\bibnamefont {Rainer}},\
  }\bibfield  {title} {\bibinfo {title} {Virial theorem for ginzburg-landau
  theories with potential applications to numerical studies of type-ii
  superconductors},\ }\href@noop {} {\bibfield  {journal} {\bibinfo  {journal}
  {Physical Review B}\ }\textbf {\bibinfo {volume} {39}},\ \bibinfo {pages}
  {9573} (\bibinfo {year} {1989})}\BibitemShut {NoStop}%
\bibitem [{\citenamefont {Babaev}\ \emph
  {et~al.}(2002{\natexlab{b}})\citenamefont {Babaev}, \citenamefont {Faddeev},\
  and\ \citenamefont {Niemi}}]{babaev2002}%
  \BibitemOpen
  \bibfield  {author} {\bibinfo {author} {\bibfnamefont {E.}~\bibnamefont
  {Babaev}}, \bibinfo {author} {\bibfnamefont {L.~D.}\ \bibnamefont
  {Faddeev}},\ and\ \bibinfo {author} {\bibfnamefont {A.~J.}\ \bibnamefont
  {Niemi}},\ }\bibfield  {title} {\bibinfo {title} {Hidden symmetry and knot
  solitons in a charged two-condensate bose system},\ }\href@noop {} {\bibfield
   {journal} {\bibinfo  {journal} {Physical Review B}\ }\textbf {\bibinfo
  {volume} {65}},\ \bibinfo {pages} {100512} (\bibinfo {year}
  {2002}{\natexlab{b}})}\BibitemShut {NoStop}%
\bibitem [{\citenamefont {Zhang}\ \emph {et~al.}(2016)\citenamefont {Zhang},
  \citenamefont {Becerra}, \citenamefont {Covaci},\ and\ \citenamefont
  {Milo\ifmmode \check{s}\else \v{s}\fi{}evi\ifmmode~\acute{c}\else
  \'{c}\fi{}}}]{zhang2016}%
  \BibitemOpen
  \bibfield  {author} {\bibinfo {author} {\bibfnamefont {L.-F.}\ \bibnamefont
  {Zhang}}, \bibinfo {author} {\bibfnamefont {V.~F.}\ \bibnamefont {Becerra}},
  \bibinfo {author} {\bibfnamefont {L.}~\bibnamefont {Covaci}},\ and\ \bibinfo
  {author} {\bibfnamefont {M.~V.}\ \bibnamefont {Milo\ifmmode \check{s}\else
  \v{s}\fi{}evi\ifmmode~\acute{c}\else \'{c}\fi{}}},\ }\bibfield  {title}
  {\bibinfo {title} {Electronic properties of emergent topological defects in
  chiral $p$-wave superconductivity},\ }\href
  {https://doi.org/10.1103/PhysRevB.94.024520} {\bibfield  {journal} {\bibinfo
  {journal} {Phys. Rev. B}\ }\textbf {\bibinfo {volume} {94}},\ \bibinfo
  {pages} {024520} (\bibinfo {year} {2016})}\BibitemShut {NoStop}%
\bibitem [{\citenamefont {Su}\ and\ \citenamefont {Lin}(2018)}]{ying2018}%
  \BibitemOpen
  \bibfield  {author} {\bibinfo {author} {\bibfnamefont {Y.}~\bibnamefont
  {Su}}\ and\ \bibinfo {author} {\bibfnamefont {S.-Z.}\ \bibnamefont {Lin}},\
  }\bibfield  {title} {\bibinfo {title} {Pairing symmetry and spontaneous
  vortex-antivortex lattice in superconducting twisted-bilayer graphene:
  Bogoliubov-de gennes approach},\ }\href
  {https://doi.org/10.1103/PhysRevB.98.195101} {\bibfield  {journal} {\bibinfo
  {journal} {Phys. Rev. B}\ }\textbf {\bibinfo {volume} {98}},\ \bibinfo
  {pages} {195101} (\bibinfo {year} {2018})}\BibitemShut {NoStop}%
\bibitem [{\citenamefont {Wu}\ and\ \citenamefont
  {Das~Sarma}(2019)}]{fengcheng2019}%
  \BibitemOpen
  \bibfield  {author} {\bibinfo {author} {\bibfnamefont {F.}~\bibnamefont
  {Wu}}\ and\ \bibinfo {author} {\bibfnamefont {S.}~\bibnamefont {Das~Sarma}},\
  }\bibfield  {title} {\bibinfo {title} {Identification of superconducting
  pairing symmetry in twisted bilayer graphene using in-plane magnetic field
  and strain},\ }\href {https://doi.org/10.1103/PhysRevB.99.220507} {\bibfield
  {journal} {\bibinfo  {journal} {Phys. Rev. B}\ }\textbf {\bibinfo {volume}
  {99}},\ \bibinfo {pages} {220507} (\bibinfo {year} {2019})}\BibitemShut
  {NoStop}%
\end{thebibliography}%

\end{document}